\documentstyle[preprint,aps]{revtex}

\newcommand{\bm}[1]{\mbox{\boldmath $#1$}}

\begin{document}
\draft
\title{Circular phase of a two dimensional ferromagnet with dipolar 
interactions} 

\author{F. Matsubara and J. Sasaki}
\address{Department of Applied Physics, Tohoku University, Sendai 980-77,
Japan}

\date{\today}

\maketitle

\begin{abstract}

We make for the first time a large-scale Monte-Carlo simulation of a 
ferromagnetic Heisenberg model with dipolar interactions on a two 
dimensional square lattice with open boundaries 
using an efficient new technique. 
We find that a phase transition occurs in the model and 
the ordered phase is characterized by a circular arrangement of the spins.

\end{abstract}

\pacs{75.10.Hk, 75.10.Nr, 75.40.Mg}

Spin structures in systems with competing interactions have been 
of great interests in the last decades. 
In the homogeneous systems, competition between the nearest neighbor 
and the second nearest neighbor interactions brings a helical spin ordering 
in the Heisenberg model\cite{Yoshimori} 
and a devil's staircase in the Ising model\cite{Bak}.  
In the random systems, spin glass phase occurs due to the competition between 
ferromagnetic and antiferromagnetic exchange interactions
\cite{Edwards,Sakata} and the competition in the anisotropy between 
two ions in the antiferromagnetic mixture causes a new magnetic ordering 
called as an oblique antiferromagnet\cite{Aharony,OAF,Katsumata}. 
In these cases, magnitudes of the competing interactions are of the same 
order and the spin structures are modulated  microscopically. 
Another kind of competition occurs in the ferromagnetic system, 
namely, the exchange interactions compete with the dipolar 
interactions which are much smaller than the formers. 
This competition has been studied in technological interest rather than the 
physical one, because it realizes magnetic domains as in the real materials. 
Magnetic structures of ferromagnetic thin films have been attracted in 
recent years, especially in the field of the magnetic storage technologies 
\cite{Argyle,Smith,Allenspach}. 
However, the models studied so far are phenomenological ones in which 
systems are divided into small cells with a certain magnetization and 
anisotropy energy and the distribution of the magnetizations is determined 
using the Landau-Lifshitz-Gilbert equation 
\cite{Yan,Nakatani,Opheusden,Mansuripur}.

Although the models reproduce the domain structures of those films fairly 
well, there still remains primary questions whether or not those structures 
could really be reproduced in a microscopic model and, if it could, 
whether or not a clear-cut phase transition could be seen? 
To answer the questions, we are necessary to make a large-scale 
Monte Carlo(MC) simulation of a Heisenberg model with dipolar interactions, 
because the domains can be seen only in large systems. 
The bottleneck of the simulation is, however, to calculate the 
dipolar fields of the individual spins which needs the CPU time of order 
$N^2$ in every MC step, where $N$ is the number of the total spins. 
Because of this reason, no large-scale MC simulation study has been done up 
to now on the spin structure as well as the phase transition of this 
model\cite{Taylor,Hucht}.

In this letter, we report for the first time results of a large-scale 
MC simulation of a ferromagnetic Heisenberg model on the two dimensional 
square lattice with open boundaries. 
We develop an efficient new MC technique which reduces 
the CPU time of order from $N^2$ to $N\log_2N$. 
We find that a clear-cut phase transition really occurs at a finite 
temperature and a domain-like spin structure appears below that temperature. 
The structure is shown to be characterized not by closure ferromagnetic 
domains but by a circular arrangement of the spins. 
We call, hence, the ordered phase as a circular phase. 
Values of the critical exponents are also estimated to examine 
the nature of the phase transition.

We start with the classical Heisenberg model described by the Hamiltonian:
\begin{equation}
 H =  - J\sum_{<i,j>}\bm{S}_i\bm{S}_j 
   + D\sum_{i\neq j}(\frac{\bm{S}_i\bm{S}_j}{r_{ij}^3} - 
      \frac{3(\bm{S}_i\bm{r}_{ij})(\bm{S}_j\bm{r}_{ij})}{r_{ij}^5}), 
\end{equation}
where $|\bm{S}_i| = 1$. 
The former term describes the exchange energy, where $J ( > 0)$ denotes the 
exchange constant and $\langle i,j \rangle$ runs over the nearest neighbor 
pairs. 
The latter term describes the dipolar energy, where $D = (g\mu_BS)^2/a^3$ but 
being considered here as a parameter 
and $r_{ij}$ is the distance between $i$th and $j$th spins 
which is measured in the units of the lattice constant $a$. 
Hereafter we measure $J$ and $D$ in the units of $k_B = 1$ with 
$k_B$ being the Boltzmann constant. 
The lattice is the square lattice of $L \times L$ with open boundaries. 
We choose the $x$- and $y$-axes being parallel to the lattice plane and the 
$z$-axis perpendicular to them.

The MC simulation is made as follows. 
Since the dipolar interactions are of long-range, huge CPU time is necessary 
to calculate the dipolar field $\bm{H}_D^{(i)}$ for every spin $\bm{S}_i$. 
Then we distinguish the dipolar field $\bm{H}_D^{(i)}$ into two 
parts  $\bm{H}_{D1}^{(i)}$ and $\bm{H}_{D2}^{(i)}$, i.e., $\bm{H}_{D1}^{(i)}$ 
is contributions of the spins up to the second nearest neighbors and 
$\bm{H}_{D2}^{(i)}$ those of the other spins. 
We calculate $\bm{H}_{D1}^{(i)}$ as well as the exchange field for 
every update trial. On the other hand, $\bm{H}_{D2}^{(i)}$ 
is updated for every $m$ MC steps(here we choose $m = 5$). 
Therefore, we can calculate the set of the dipole fields 
\{$\bm{H}_{D2}^{(i)}$\} simultaneously by using FFT algorithm which needs 
the CPU time of order $N\log_2N$. 
Of course, errors arise which will become larger as the temperature is 
decreased.  However, we think that the errors are not so serious 
in the thermal equilibrium because of the following reasons. 
Since a large number of the spins may not change their directions 
in each MC step, the change of the dipolar field $\bm{H}_{D2}^{(i)}$ 
due to the changes of far spins will not be accumulated considerably. 
Then the major part of the errors will come from the ignorance of the updates 
of near spins, especially those on the third nearest neighbor lattice sites. 
Therefore, the errors will become considerable below $T \sim 4 \times D/2^3 
= D/2$. 
However, at such low temperatures, the spins will change their directions 
very little in each MC step. 
Hence we believe our method does not change results significantly at all 
temperatures\cite{Sasaki}.

We carry out the simulation of the model with $J = 1$ and $D = 0.1$ 
using the heat bath method\cite{heat}. 
The linear sizes of the lattice are $L = 16 - 256$.  
We make the MC simulation for several times starting from different 
initial spin configurations. 
The MC steps are about 100,000 for the largest lattice. 
Cooling the temperature, we calculate the specific heat $C$, 
the absolute value of the uniform magnetization $M_u$ and the susceptibility 
$\chi^{\mu} \;(\mu = x, y, z)$: 
\begin{eqnarray}
  C   &=& \frac{<H^2> - <H>^2}{T^2N}, \\
  M_u &=& <|\bm{M}|>,                 \\
  \chi^{\mu} &=& \frac{<(M^{\mu})^2>N}{T}, 
\end{eqnarray}
where $\bm{M}=(1/N)\sum_i\bm{S}_i$, and $<...>$ means the thermal 
average. We also observe  the spin structure.

In Fig. 1, we show typical three spin structures A, B and C observed 
at low temperatures. The structure A is single-domained and the structure C 
like domains of closure which frequently observed in thin films\cite{Argyle}. 
The structure A is seen for smaller lattices of $L \leq 16$ 
and the structure C for larger lattices of $L \geq 32$. 
The structure B is an incomplete one which occurs for 
intermediate size of $L \sim 24$. 
This relation  between the spin structure and the lattice size is not 
so clear-cut.  For $L = 24$, the structure B dominates over the structures 
A and C, but for $L = 32$ the structure C is observed most frequently.

We show the specific heat $C$ and the z-component of the susceptibility 
$\chi^z$ in Figs. 2 and 3, respectively. 
The specific heat exhibits a size dependent sharp peak at $T \sim 0.85$. 
Hereafter, we denote the peak temperature as $T_c$. 
The susceptibility also exhibits a size dependent maximum at almost 
the same temperature of $T_c$. These results suggest the occurrence of 
some long-range order characterized by the xy-components of the spins. 
If the long-range order really occurs, it would be one with 
the spin structure C shown in Fig.1, because we could not observe any 
evidence of another phase transition below $T_c$. 
Then we calculate, in addition to $M_u$, the absolute value of 
the circular component of the magnetization $M_{\phi}$: 
\begin{eqnarray}
 M_{\phi} = <|[\frac{1}{N}\sum_i(\bm{S}_i \times \frac{\bm{r}_i-\bm{r}_c} 
 {|\bm{r}_i-\bm{r}_c|})]_z|> ,
\end{eqnarray}   
where $\bm{r}_i$ and $\bm{r}_c$ are the position vectors of the $i$th spin 
and the center of the lattice, respectively.
In Fig.4, we show temperature dependences of $M_u$ for different $L$.  
When $L$ is small, $M_u$ increases as the temperature is decreased and seems 
to reach 1 at $T = 0$. 
However, as $L$ is increased, $M_u$ only exhibits a hump around $T_c$.  
In Fig.5, we show temperature dependences of $M_{\phi}$ for different $L$. 
We see an opposite size dependence, i.e., $M_{\phi}$ for smaller $L$ 
exhibits a hump around $T_c$, while $M_{\phi}$ for larger $L$ exhibits a 
rapid increase below $T_c$.  
Moreover, the size dependence of $M_{\phi}$ is reversed as the temperature 
is decreased below $T_c$. That is, $M_{\phi}$ for a larger $L$ has a larger 
value below $T_c$. 
These results clearly reveal that the long-range order really occurs 
and it is described not by the uniform magnetization but 
by the circular component of the magnetization.

We should mention that the ordered phase obtained above is not the one 
composed of four ferromagnetic domains with their magnetization axes 
parallel to the edges of the lattice, because if so, 
one would get $M_{\phi} \sim \sinh^{-1}(1) \sim 0.88$ at $T = 0$, 
whereas the result in Fig. 5 suggests $M_{\phi} \sim 1$ at $T = 0$. 
Moreover, we could not see any domain-like structure near below $T_c$. 
Therefore it is natural to call the phase as a circular phase 
rather than the ferromagnetic phase with domains. 
Next we examine the phase transition of the model using the 
finite-size-scaling analysis. If the second order phase transition occurs 
at $T = T_c$, $M_{\phi}$ for the lattice 
with $L$ will be scaled as
\begin{eqnarray}
 M_{\phi}L^{\beta / \nu} = F(tL^{1/\nu}),
\end{eqnarray}
where $t = (T-T_c)/T_c$, $\beta$ and $\nu$ are exponents of 
the order parameter $M_{\phi}$ and the correlation length, and $F$ is some 
scaling function\cite{Barber}. 
We could fit the data for $L \geq 64$ fairly well, which is shown in Fig. 6. 
From the results, we estimate values of the exponents as $\beta \sim 0.23$ 
and $\nu \sim 1.20$ together with $T_c \sim 0.88$. 
We have also estimated the values of $\beta$ and $\alpha$  by plotting 
$M_{\phi}$ vs $T$ and $C$ vs $T$ in log-log forms, respectively, and obtained 
similar values of $\beta \sim 0.2$, $\alpha \sim -0.4$ and $T_c \sim 0.87$. 
Note that this value of $\alpha$ is compatible with the scaling 
relation of $\alpha = 2 - d\nu$ with $d = 2$. 
These values are not similar neither to those of the Ising model of 
$\beta = 1/8$ and $\nu = 1$ nor to those of an anisotropic xy-like model 
without dipolar interactions\cite{Sasaki2}. 
We conclude, hence, that a usual second order phase transition really occurs 
in this model and it belongs to a new universality class. 
This is another evidence that the circular phase is different from the 
ferromagnetic phase with domains. 
We note here that the present result is not incompatible with experimental 
observations. 
If an axial anisotropy along the edges of the lattice exists, 
the spins tend to align the directions parallel to the edges. 
Thus the circular structure is deformed into the one with four ferromagnetic 
domains of closure which are observed in thin films\cite{Argyle}. 
Finally we should emphasize that the results presented here have been 
obtained by using the new MC technique within the CPU time much less than 
that of the conventional MC method.

The authors would like to give their thank to Dr. K. Tan for leading them 
to this problem, and to Professor T. Shirakura and Dr. T. Nakamura for their 
valuable discussions. 
This work was financed by a Grant-in-Aid for Scientific Research from the 
Ministry of Education, Science and Culture.

\begin{figure} 
\caption{Snapshots of the spin structures for different sizes of the lattice 
at $T = 0.1$. }
\end{figure}

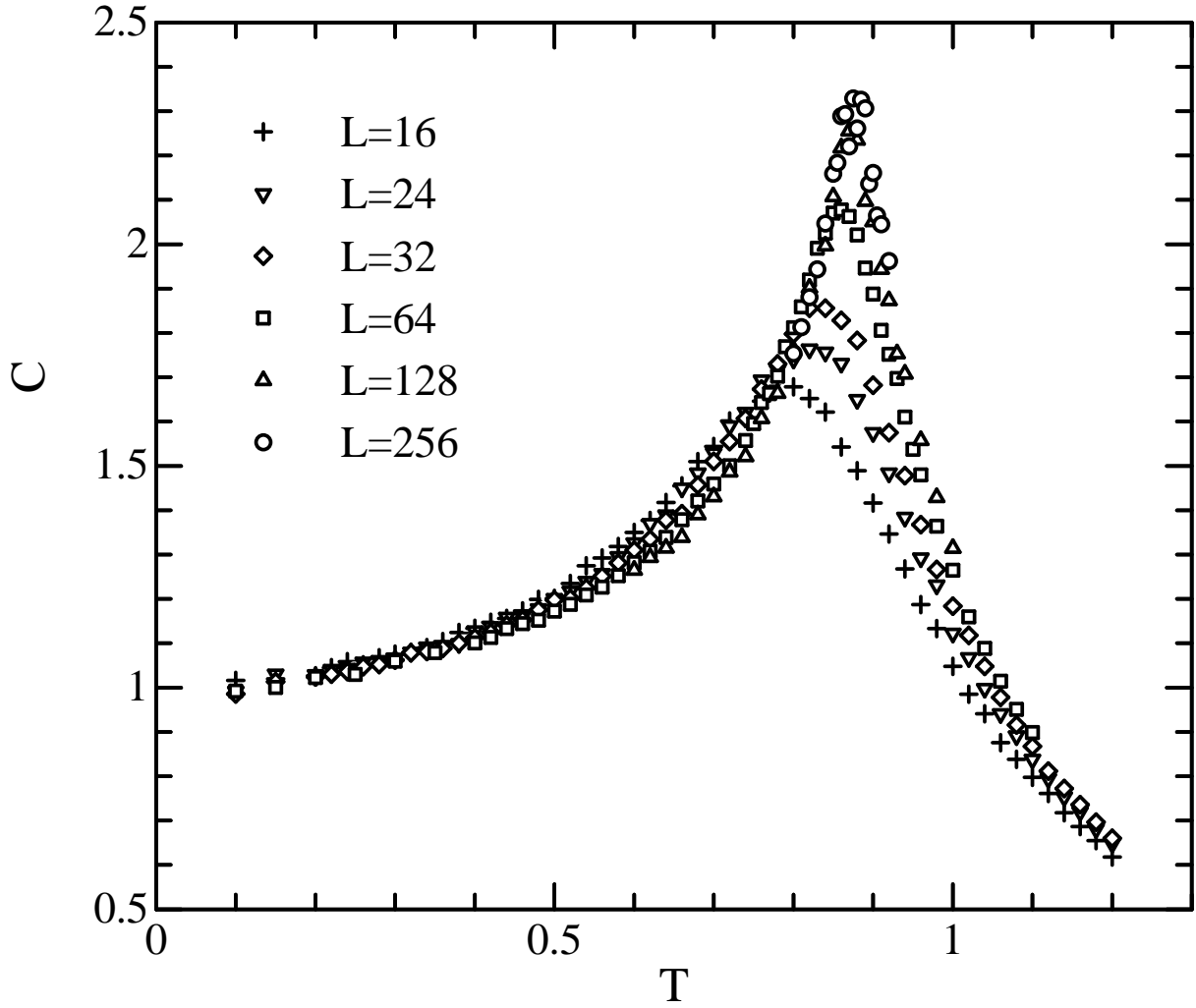
\begin{figure} 
\caption{
Temperature dependences of the specific heat $C$ for different sizes 
of the lattice. }
\end{figure}

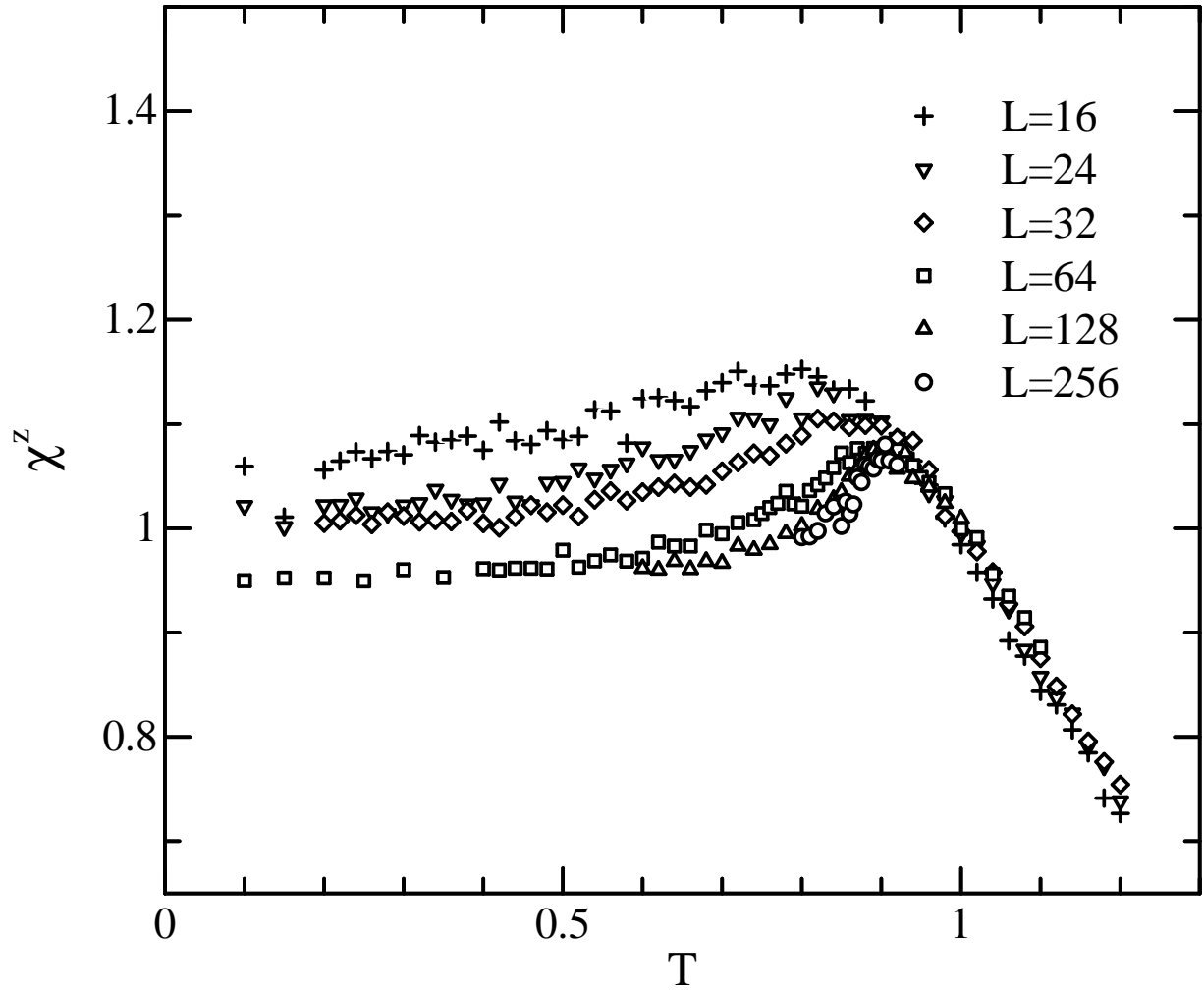
\begin{figure} 
\caption{
Temperature dependences of the susceptibility $\chi^z$ for different sizes 
of the lattice. }
\end{figure}

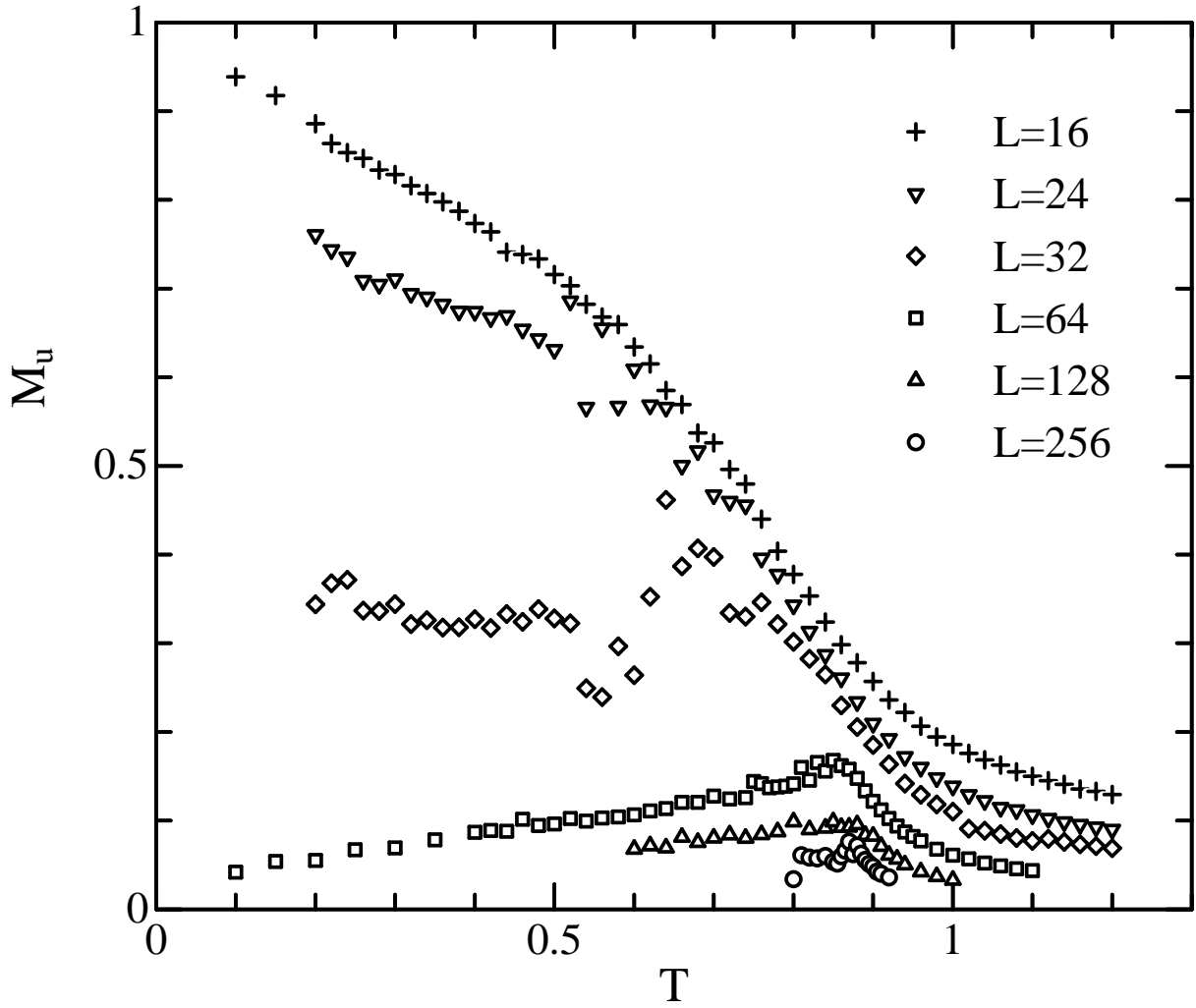
\begin{figure} 
\caption{
Temperature dependences of the absolute magnetization $M_u$ for different sizes 
of the lattice. }
\end{figure}

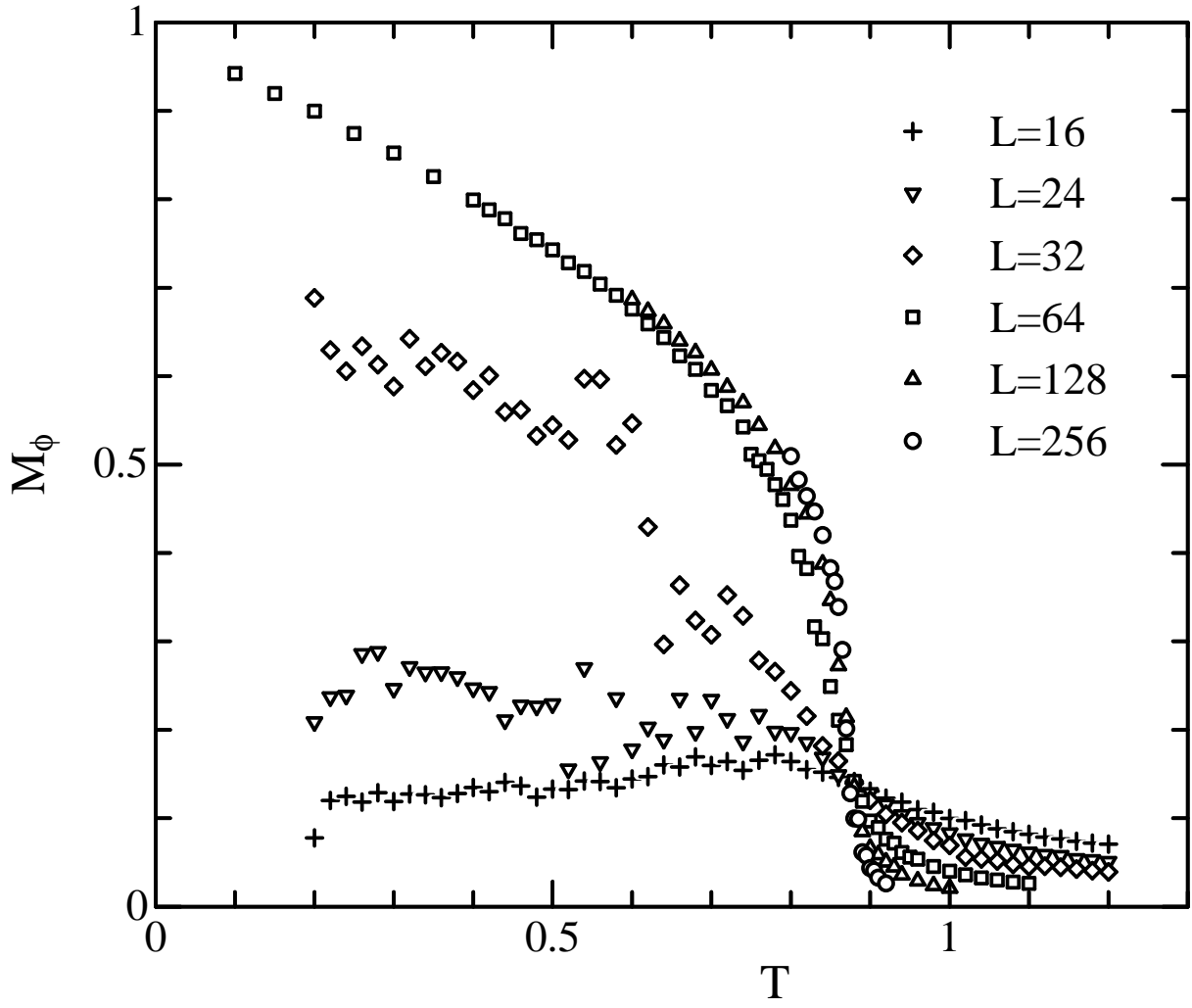
\begin{figure} 
\caption{
Temperature dependences of the circular component of the magnetization 
$M_{\phi}$ for different sizes of the lattice. }
\end{figure}

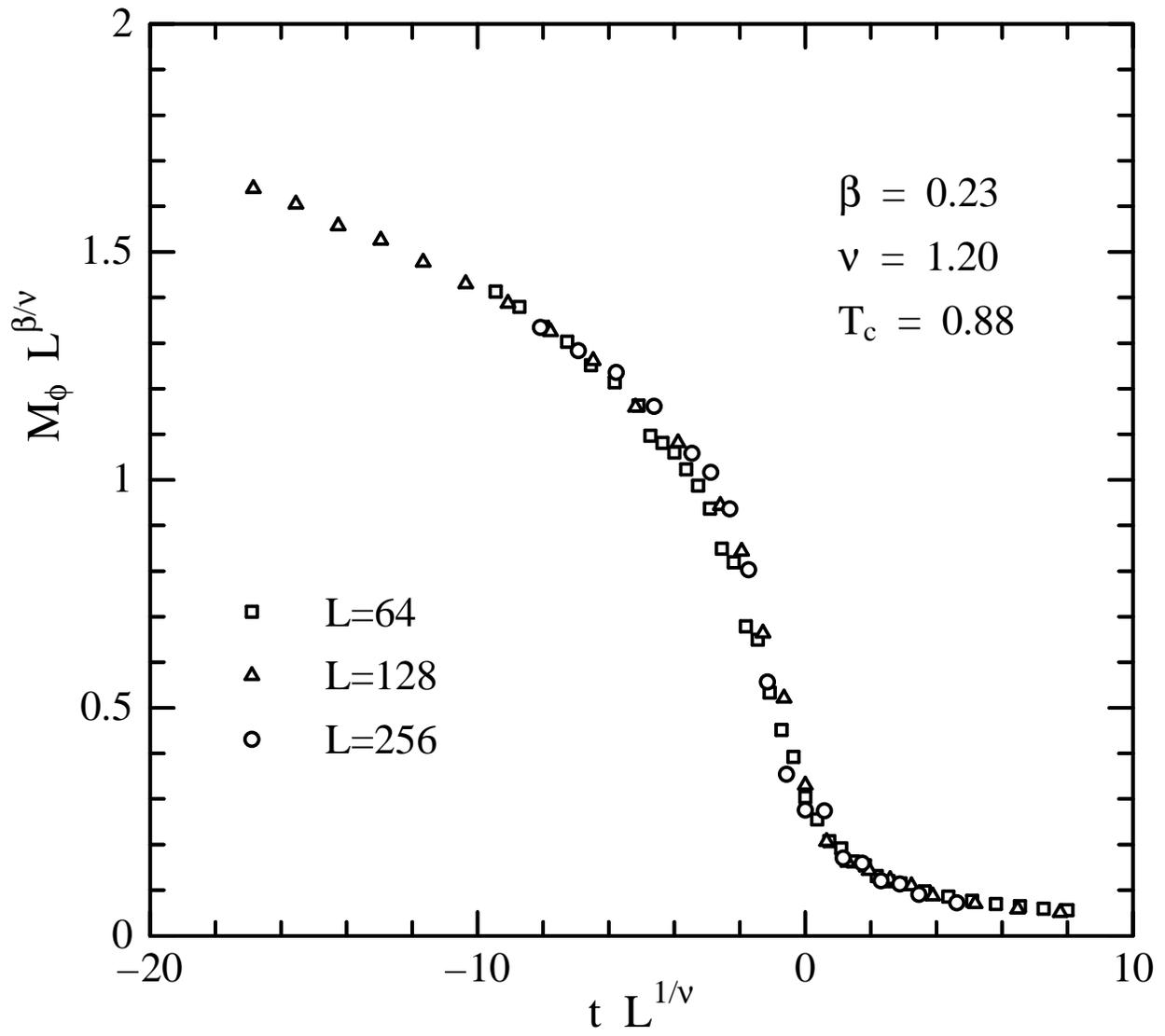
\begin{figure} 
\caption{ A finite-size-scaling plot of $M_{\phi}$. }
\end{figure}


\begin{references}
\bibitem{Sasaki} We have also made the simulation of the model for smaller 
lattices of $L = 16$ and $24$ using a conventional MC method and found that 
our method reproduces all the results satisfactorily except for a slight 
reduction of the specific heat for $T < T_c$, which will be published 
elsewhere.
\bibitem{Sasaki2} We have also made the simulation of an anisotropic 
Heisenberg model with $J^x = 1.0$, $J^y = 0.9$ and $J^z = 0.8$ and obtained 
$\beta \sim 0.19$ and $\nu \sim 0.8$, which will be published elsewhere.
\end{references}
\end{document}